\documentclass{article}
\usepackage{spconf, graphicx,graphics, epsfig, cite, url} 
\usepackage{amssymb, amsmath, amsbsy, amsfonts, algorithmic, textcomp, mathptmx, time, mathtools}
\usepackage{setspace}
\title{Seizure Detection using Least EEG Channels by \\ 
Deep Convolutional Neural Network}

\name{
Mustafa Talha Avcu$^1$, 
Zhuo Zhang$^2$, 
Derrick Wei Shih Chan$^3$ 
\address{
\small$^1$Electrical \& Electronics Engineering, Bilkent University, Turkey\\ 
\small $^2$Neural \& Biomedical Engineering, Institute for Infocomm Research, Singapore \\ 
\small $^3$Dept of Paediatrics, Neurology Service, KK Women's and Children's Hospital, Singapore
}}

\begin{document}

\maketitle
\thispagestyle{empty}
\pagestyle{empty}

\begin{abstract}

This work aims to develop an end-to-end solution for seizure onset detection. We design the SeizNet, a Convolutional Neural Network for seizure detection. To compare SeizNet with traditional machine learning approach, a baseline classifier is implemented using spectrum band power features with Support Vector Machines (BPsvm). We explore the possibility to use the least number of channels for accurate seizure detection by evaluating SeizNet and BPsvm approaches using all channels and two channels settings respectively. EEG Data is acquired from 29 pediatric patients admitted to KK Woman's and Children's Hospital who were diagnosed as typical absence seizures. We conduct leave-one-out cross validation for all subjects. Using full channel data, BPsvm yields a sensitivity of 86.6\% and 0.84 false alarm (per hour) while SeizNet yields overall sensitivity of 95.8 \% with 0.17 false alarm. More interestingly, two channels seizNet outperforms full channel BPsvm with a sensitivity of 93.3\% and 0.58 false alarm. We further investigate interpretability of SeizNet by decoding the filters learned along convolutional layers. Seizure-like characteristics can be clearly observed in the filters from third and forth convolutional layers. 

\let\thefootnote\relax\footnotetext{This work has been submitted to the IEEE for possible publication. Copyright may be transferred without notice, after which this version may no longer be accessible.}

\end{abstract}

\section{INTRODUCTION}

Monitoring brain activity through EEG is critical for epilepsy diagnosis. To capture seizure events that may occur sparsely, neurologists have to visually scan vast amount of EEG data. The process is extremely time consuming and may be subjective due to inter observer variance. Computer aided seizure detection approach would serve as valuable clinical tool for the scrutiny of EEG data in an objective and much more efficient manner. 

Traditional machine learning approaches for seizure detection usually composite three stages: data pre-processing to eliminate artifacts, followed by feature extraction and decision-making. A number of features have been identified to describe the behavior of seizures, including those based on time-domain, frequency-domain, time-frequency analysis, wavelet features and chaotic features such as entropy etc. A pioneer work was presented in \cite{Shoeb}, which created subject-specific seizure onset detection model using hand-crafted features extracted from the raw EEG data followed by classification. The subject-specific model reaches a sensitivity of 96\% and false alarm rate of 0.08 per hour on CHB-MIT dataset by using SVM over a combination of spectral, spatial and temporal features. 

The extracted features are believed to contain discriminative information for systems to differentiate seizure from non-seizure states. Feature space are highly compact as compare to raw EEG data space which is critical when computing power is limited. Furthermore, the features may bring interpretability for a machine learning system. However, there are limitations for such seizure detection methods.  Firstly, extracting features from raw data may induce information loss. Secondly, the standard power band analysis which splitting the spectrum bands (delta 0-4Hz, theta 4-8Hz etc.) could not take into consideration of individual variance of spectrum distributions. Finally, hand crafted feature extraction brings extra computational complexity for real-time applications. 

Deep learning (DL) solves such problems by representation learning which enables computer to learn high level features from raw data without human interference. Recently advance in DL such as batch normalization \cite{batchnorm}, dropout \cite{dropout} and various new network structures have largely prompt the applications for DL in real world problems and some have achieved near human-like performance. Various DL approaches have been proposed for seizure detection. Hugle \emph{et. al.} \cite{Hugle} proposed a Convolutional neural network (CNN) designed for implantable microcontroller by using only 4 electrodes selected a priori by expert. Ullah and colleagues \cite{Ullah} developed a pyramid CNN for Epilepsy Detection. A typical 13-layer CNN model is created for seizure detection problem in \cite{Acharya} using Bonn University database \cite{Anrzejak} which has only 3.27 hours of EEG data. The deep CNN structure has about 100k parameters, however neither dropout nor batch normalization is used to regularize the deep CNN. As a result, they could not reach the state-of-art performance achieved by machine learning approaches using hand crafted features. 

In this work, we aim to develop an end-to-end solution for seizure onset detection. A CNN structure called seizNet is carefully designed to enable an efficient and effective representative learning for seizure onset detection, equipped with dropout and batch normalization to prevent overfitting for a more generalized solution. We explore the possibility to use least number of channels for accurate detection across subjects. Finally, we attempt to interpret the model by discovering signatures hidden in the filters from different convolutional layers.

\section{METHOD}

\subsection{Baseline method -- BPsvm}
We develop a SVM-based classifier using hand crafted spectrum band power features, called BPsvm. 
Frequency spectrum components within the 0-25 Hz band is considered, as suggested by \cite{Gotman}. In this study, we  preserve the 5-second epoch for analysis across different approaches. To obtain features in higher resolution, we split the 5-second epoch into 5 1-second windows for spectrum transformation and band power feature extraction. In Shoeb's work
\cite{Shoeb}, 8 bands from 0.5-24 Hz are chosen. As our window size is 1 second, the least sampling rate is 1 Hz, our sub-bands are defined as [1-3, 3-6, 6-9, 9-12, 12-15, 15-18, 18-21, 21-24] Hz. Spectrum band power feature in the
sub-band signals on every 1-second are calculated and then are concatenated into one instance for every 5-second epoch. SVM
classifier is trained using radial basis function (RBF) kernel. 

\subsection{Deep learning method -- SeizNet}
We develop a deep CNN network named SeizNet for end-to-end seizure detection solution. Comparing to 
\cite{Acharya}, SeizNet contains additional dropout layers \cite{dropout} and batch normalization 
\cite{batchnorm} after every convolution layer. Such layers are designed to avoid model overfitting. 
Unlike the typical usage of dropout which is only after fully connected layer, we used dropout in 
various parts of the model as it is indeed suggested by the inventors of dropout \cite{dropout}. 
The number of filter at each convolution layer is multiplied by two every time like VGGNet \cite{Simonyan}. 
It enables SeizNet to have less number of filters at low levels in which filters learn basic shapes, 
while having more filters at the higher levels where filters are capable of grasping sophisticated patterns. 
As an activation function ReLU is used and other hyper-parameters of the model such as number of filters and filter sizes at each layer as well as number of unit in fully connected layer are cross-validated over a broad range. 
Detailed architecture of SeizNet can be found in table \ref{SeizNet}. The total number of parameters 
for SeizNet-2chn and SeizNet-18chn are $200,592$ and $201,872$ respectively, both include $240$ non-trainable parameters. 

\begin{table}[ht]
\footnotesize
\caption{SeizNet Architecture}
\label{SeizNet}
\begin{center}
\begin{tabular}{l|l|c}
\hline
& Layer & Output\\
\hline
\hline
Input           & ($1000 \times n^*$) & ($1 \times 1000 \times n$)\\
\hline
Conv 1 &  $8 \times Conv2D(1 \times 10)$ & $(1 \times 991 \times 8)$\\
& MaxPool2D $(1\times2)$   & $(1\times495\times8)$\\
& Dropout(0.2)	 & $(1\times495\times8)$\\
\hline
Conv 2 & $16 \times Conv2D(1\times10)$ & $(1\times486\times16)$\\
& MaxPool2D $(1\times2)$   & $(1\times243\times16)$\\
& Dropout (0.2)	 & $(1\times243\times16)$ \\
\hline
Conv 3 & $32 \times Conv2D(1 \times10)$ & $(1\times 224 \times32)$\\
& MaxPool2D $(1\times2)$   & $(1 \times 112\times 32)$\\
& Dropout (0.2)	 & $(1\times112 \times 32)$\\
\hline
Conv 4 & $64 \times Conv2D(1\times10)$ & $(1\times93 \times64)$\\
& MaxPool2D ($1 \times 2$)   & $(1\times46 \times64)$\\
& Dropout (0.2)	 & $(1 \times46 \times64)$\\
\hline
Flatten &  Flatten			& (2944)\\
& Dense (50)		& 50\\
 & Dropout (0.5)	& (50)\\
\hline
Output & Dense (2)		& (2)\\
\hline
\multicolumn{3}{r}{\footnotesize{*n=2 for 2-channel data, n=18 for 18-channel data}} 
\end{tabular}
\end{center}
\end{table}

We conduct Leave-one-subject-out cross validation to evaluate our model. For training, we use Adam \cite{Adam} optimizer with a learning rate of $4.1e-3$, binary cross entropy loss function and batch size of 128. We explored early-stopping \cite{Yao} approach by randomly selecting 20\% of training data as validation split. However, it is observed that SeizNet does not overfit due to effect of aggressive usage of dropout and batch normalization applied after every convolution layer. Therefore, 100 iterations have been decided to execute without any validation split. It also enables us to take advantage of all the data which possibly affects performance in a statistically important level since deep learning approaches for BCI problems are often deprived of big datasets.  

\subsection{Filter Decoding for SeizNet interpretation}
CNNs extract spatial features hierarchically in a modular way throughout convolution layers, the filters decompose the input space to set a mapping between abstract features and labels. Decoding such filters help us discern what decomposed components are, thereby how CNNs work. One technique to fathom the characterization of hidden units is visualizing sample inputs that maximize the selected units. A pioneer method, Activation Maximization (AM) proposed by Erhan in 2009 \cite{Erhan} has turned this technique into an optimization problem yielding artificial inputs that maximally activate any chosen hidden unit/units by gradient ascent rather than selecting from the data set which is shown to be problematic and inadequate in the sense of leading a conclusion. Furthermore, its consistency has been shown with different initializations producing mostly same salient features at the input\cite{Erhan}. 

To date, AM has been widely used to decode abstract spatial filters and to make qualitative interpretations of CNNs. It was first applied by \cite{Simonyan14} to the AlexNet \cite{alexNet}, known as first modern CNN architecture as well as to the others such as VGGNet \cite{Simonyan}, GoogleNet \cite{Szegedy} etc.

\section{Experiment and Result}

\subsection{Seizure EEG data and Pre-processing} 
Data used in this study is from KK Women's and Children's Hospital, Singapore. IRB was acquired from the hospital review board. EEG data of 29 pediatric patients diagnosed with typical absence seizures are included in this study. The data are extracted from Nikon Kohden EEG-1200K and EEG-9100K recording systems  (reading setting: Cal Voltage=$50 \mu V$, HFF=$70Hz$, LFF=$0.53Hz$, Sensitivity=$7-10\mu V/mm$, sampling rate=$200 \,\, or \,\, 500Hz$). The length of patients' EEG recordings range from 25 to 66 minutes. In total the data contains 1037.6 minutes of EEG recording with 24.95 minutes seizure data distributed among 120 seizure onsets. EEG data is down-sampled to 200 Hz across all subjects. Data from all channels for each subject is z-normalized. Window size of 5 second is chosen and preserved for all the methods in order to obtain a conclusive comparison based on performance metrics. A common problem in CNN networks for seizure detection is that datasets are often imbalanced meaning interictal phases outnumber the ictal phases by a wide margin and it has been shown that imbalanced datasets lead to statistically significant performance drop in CNN architectures \cite{Buda}. To overcome this issue, a data augmentation method during pre-preprocessing is preferred rather than undersampling or oversampling. To increase the number ictal phases, sliding is applied with different overlapping proportions according to existence or absence of seizure. Namely, while shifting with 5 seconds (no overlapping) is implemented to create interictal class, 0.075 second shifting is used for ictal class to create balanced input for the SeizNet.  For BPsvm, however, no such technique is applied since SVM is shown to be robust against imbalanced datasets \cite{Akbani}. 

\subsection{Experiment Settings and Performance Metrics}
We compared result of four experimental settings including 18-channel SVM, 18-channel CNN, 2-channel SVM and 2-channel CNN respectively. Performance of the seizure detection algorithms are assessed with sensitivity and false alarm rate by the community \cite{Tzallas} and often extended with ‘latency’ in order to make more comprehensive analysis across detector algorithms. The definitions are described as follows: 

\begin{footnotesize}
\textbf{Sensitivity(\%)}: Proportion of seizures correctly detected

\textbf{False alarm rate(fp/h)}: Number of false positive seizures per hour

\textbf{Latency(second)}: Delay between electrographic onset and detection
\end{footnotesize}

While BPsvm yields determinate results, seizNet models produce different result in every round, due to randomly parameter seeding. To evaluate the result objectively, ten tests have been carried
out for both seizNet-2-chn and seizNet-18-chn models, and most frequent result which is statistically corresponds the mode of sensitivity and false alarm for a subject is chosen
as a final result.

\subsection{Results}
Results of performance metrics obtained from different experimental settings can be found in Table \ref{result}. In both BPsvm and seizNet, models trained using 18-chn reduce false alarms with a boost in sensitivity as compare to the 2-chn models. But to our surprise, SeizNet-2-chn model though using much less number of channels, outperforms BPsvm-18-chn model for all performance metrics measured. 

\begin{table}[ht]
\begin{center}
\caption{Comparison of Performance Metrics }
\label{result}
\footnotesize
\begin{tabular}{ l | c | c| c | c  } 
\hline  
Model& \multicolumn{2}{c|}{BPsvm} &  \multicolumn{2}{|c}{SeizNet} \\
\hline
Channel used           &  2-chn   &   18-chn &  2-chn   &  18-chn  \\
\hline
\hline
Seizure detected       &  104/120 & 108/120  & 112/120  &  115/120 \\
Sensitivity (\%)       &  86.6\%  &  90\%    & 93.3\%   &  95.8\%  \\
False alarms           &  33      &  14      &  10      &   3      \\ 
$FAR^*$ (fp/h)         &  1.91    &  0.81    &  0.58    &  0.17    \\
Mean Latency(sec)      &  4.42    &  3.75    &  3.26    &  3.80 \\
\hline
\multicolumn{5}{r}{\footnotesize{*FAR--false alarm rate}} 
 \end{tabular}
\end{center}
\end{table}

\begin{figure}[ht]
\centering  
\includegraphics[width=\columnwidth]{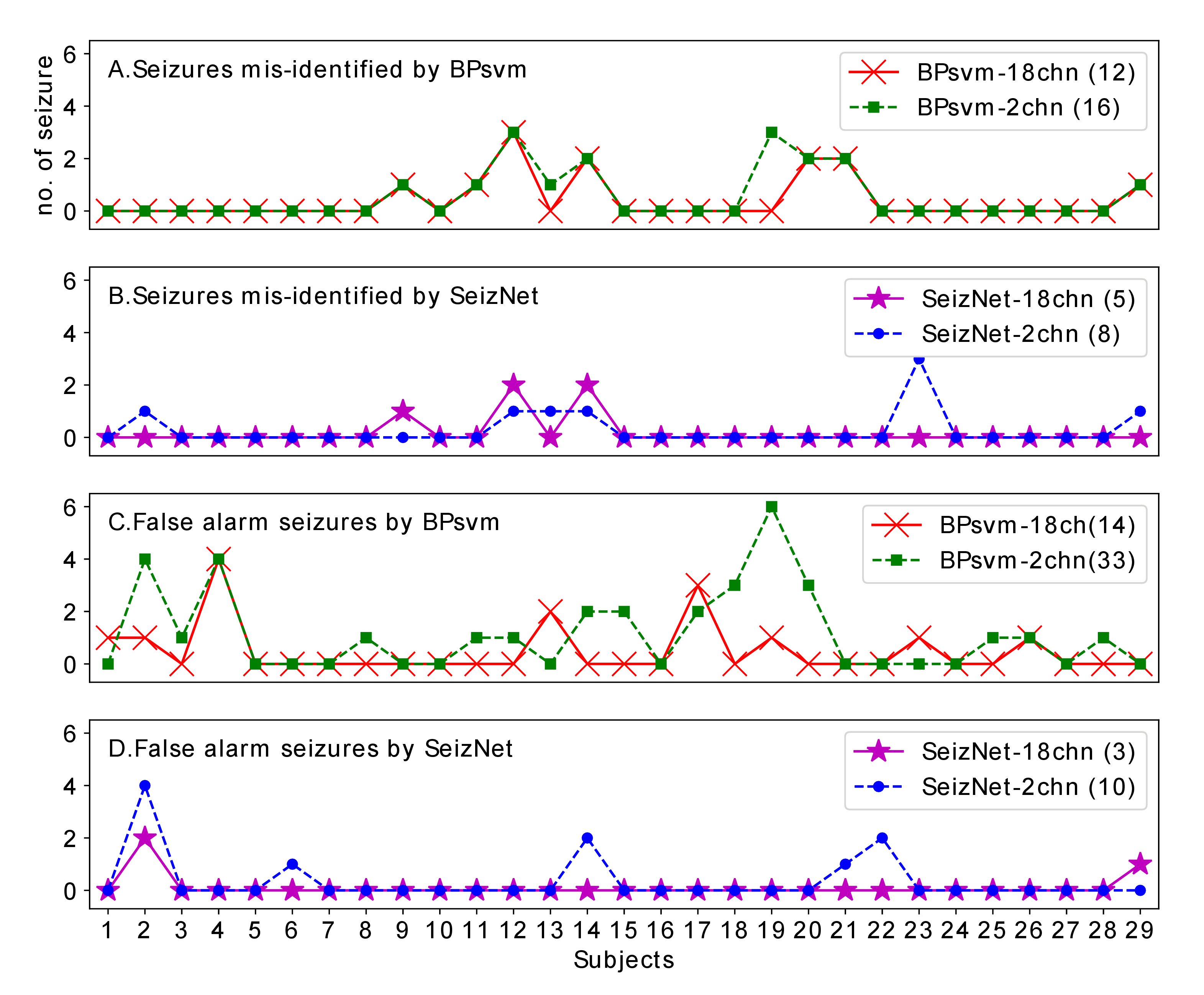}
\caption{Mis-identified and false alarm seizures across subjects}

\label{subjects}
\end{figure}

Detailed number of seizures identified correctly and false alarm for each subject can be found in Fig \ref{subjects}. 

\subsection{Filters in SeizNet}
To understand what representation features are learnt in seizNet, we use Activation Maximization method that generates input patterns activating given filters. It enables us to visualize the low level and high level features that help to reveal feature hierarchy throughout convolution layers. In this work, AM is implemented with keras-vis library \cite{kerasVis} which provides methods to generate inputs activating the given unit/units maximally. Unlike the baseline method, activation maximization loss calculated based on given input is not only dependent of model weights. In the library, there are two kind of regularization terms namely LP norm and total variation added to the loss in order to enforce natural image prior \cite{kerasVis}. Default values are preserved for the weights of these regularization terms. Only input range is changed according to our pre-processing, and it is set to (-10,10). Finally, seed for the optimization is initialized with random values.

\begin{figure}[ht]
  \centering
    \includegraphics[width=0.9\columnwidth,clip=false]{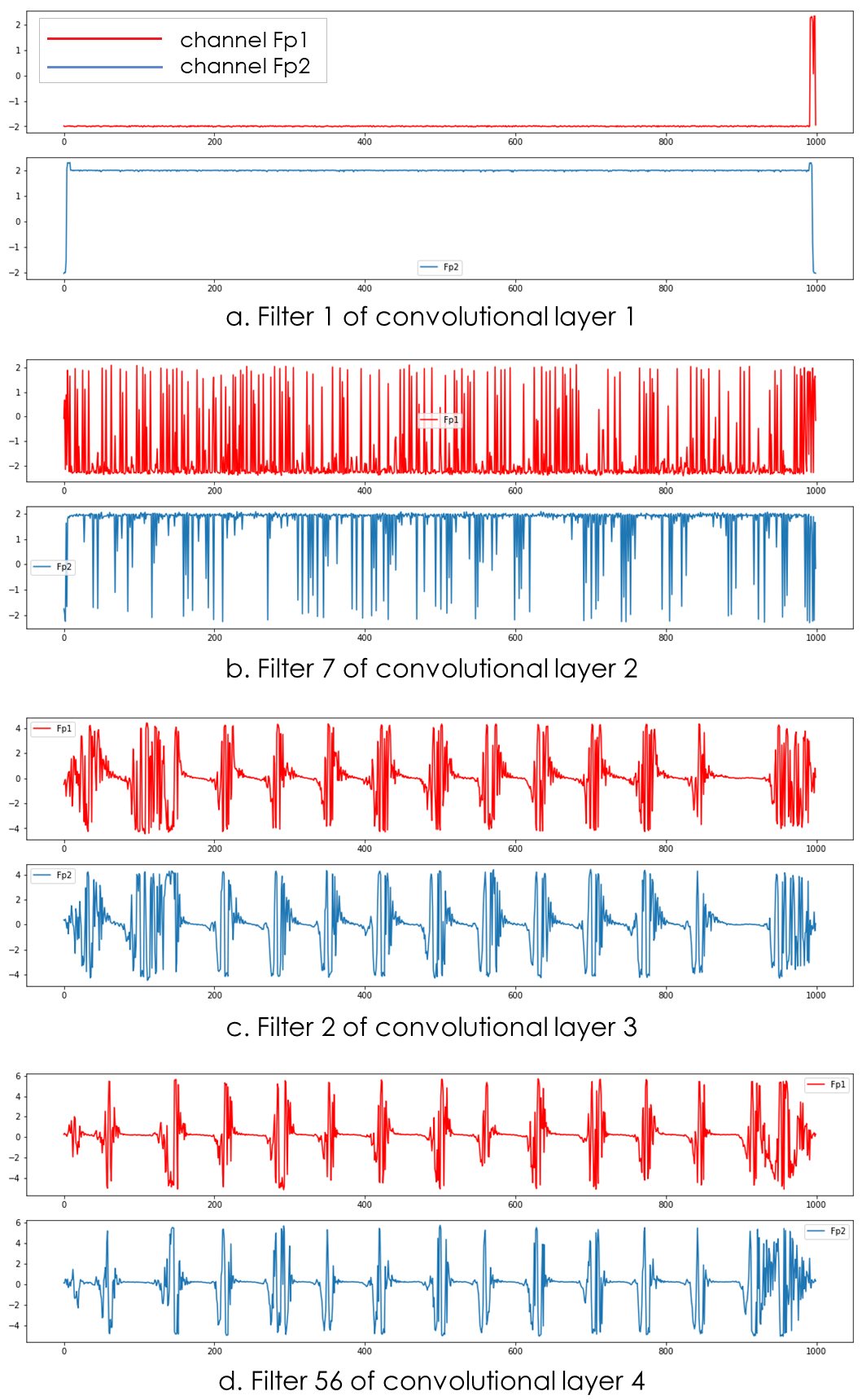}
      \caption{Filters from 4 convolutional layers}
			\setlength{\belowcaptionskip}{-1em}
			\label{vis0}
\end{figure}	
	
In image classification problems, filters at the first convolutional layer usually encode the direction and color channels. Hierarchically moving along the CNN, more and more complex features are found at the higher layers that are indeed combination of features at the lower levels \cite{Erhan}. 

In SeizNet, the first layer filters present very basic shapes as shown in Fig \ref{vis0}.a. While interpreted as color channels in the perspective of image analysis, for SeizNet the filters can be thought to encode EEG channel information, due to the fact that there are filters that have high and constant value for the channel 1 and low and constant value for the channel 2 and vice versa. Another possible interpretation could come from EEG montage perspective, where bi-polar signal represents the difference of two electrodes since the filters indeed subtract the channels one  another. 

\begin{figure}[ht]
\centering
      \includegraphics[width=\columnwidth]{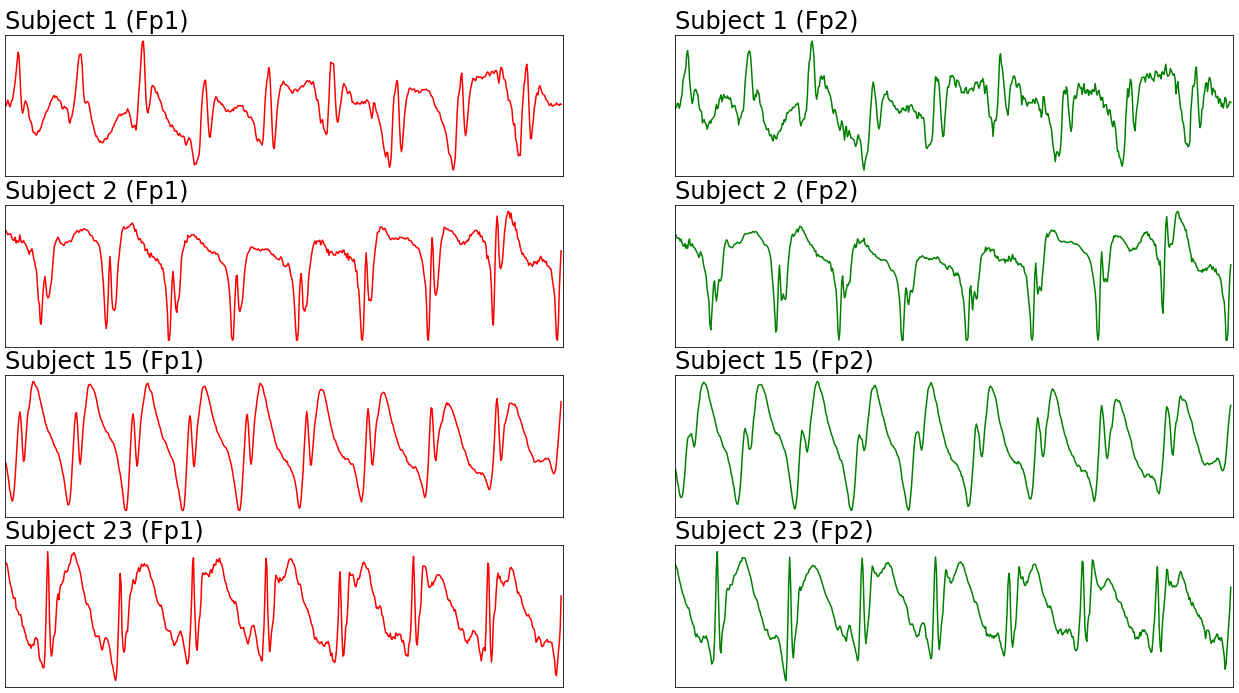}      
    \caption{Examples of seizure waveforms (3 seconds)}
		\setlength{\belowcaptionskip}{-1em}
		\label{seizure}
\end{figure}

In the second convolution layer, various kind of filters have been found but nothing meaningful to us has been observed. However, from the third convolution layer onwards we observe substantial characteristics of seizures, therefore our hypothesis for the second layer is that it serves as the middle man to bring basic information in the lower layers into  complex like-seizure signals in the higher layers

Characteristics of the seizures are observed at the third convolution layer and become clearer at the last convolution layer. It can be inferred that SeizNet has learnt the fact that absence seizures create periodic and 3 Hz signals. Nonetheless, clearly SeizNet focuses on spike-and-wave happening three times in one second and try to capture it rather than capturing a whole shape of a seizure. One possible reason behind is that seizure patterns often vary from subject to subject, as can be seen from the Fig \ref{seizure}.

\section{Discussion and Conclusion}
It has been observed from the overall results that SeizNet is better than BPsvm in terms of sensitivity and false alarm. It shows that CNN models are more suitable for generalized models unlike the SVM which is indeed often implemented as a subject-specific model. Nevertheless, it should be noted that for subject 2 and subject 23, BPsvm-2chn could identify all the seizures while not triggering more false alarm than seizNet-2chn which could not find all the seizures for the mentioned subjects. It can be inferred that for some subjects frequency domain features can be more discriminative than the features extracted from time domain. It also consistent with the fact that even EEG experts are to check time-frequency graphic in some cases in order to finalize their decision. An interesting discovery in the study is that, developed using CNN, seizNet model trained by only 2 channels' data is able to outperform traditional approach trained with full scalp EEG data. 

End-to-end approach is more favorable for developing real time seizure detection systems as it eliminates feature extraction, which can be a burden for real time signal processing. A solution using data merely from 2 channels makes the approach even more adoptable for light-weight, home based seizure monitoring system. 

\bibliographystyle{IEEEbib}
\bibliography{new_ref}
\end{document}